# Summarizing Online Conversation of Indonesia Tourism Industry using Network Text Analysis

**Andry Alamsyah, Sheila Shafira, Muhamad Alfin Yudhistira**

School of Economics and Business, Telkom University

Bandung, Indonesia

andrya@telkomuniversity.ac.id

## Abstract

Tourism Industry is one of potential revenue and has an important role in economics in Indonesia. Tourism Industry brings job and business opportunity, foreign exchange earnings, and infrastructure development, tourism also plays the role as one of the main drivers in socio-economic progress in Indonesia. The number of foreign tourists visiting Indonesia increase cumulatively and has reached 10.41 million visits or an increase of 10.46 percent from the same period in previous year. Government trying to increase the number of tourists to visit Indonesia by promoting many Indonesia tourist attraction.

To support the government effort, we need a way to understand how people's perception about tourism aspect in Indonesia. The easiest and cheapest way to see that is by extracting opinion from user generated reviews in a form of conversation among the users in social media such as Twitter. This paper proposes a method for extracting and summarizing of opinion or perception expressed by social media users. Our methods based on frequently appeared words and words relations among those dominant words. We call this method as Network Text Analysis, which is based on Social Network Analysis methodology.

As a case study, we conduct experiment against two tourism object aspect in Indonesia: Indoor and Outdoor tourist object. Specifically extracting user opinion regarding museum and nature destination from Twitter. The proposed methodology classifies topics from opinion data. Our method is fast and significantly accurate to summarize dominant topics in tourism industry when implemented in large-scale data.

## 1. Introduction

Today, Southeast Asia has entered the era of ASEAN Economic Community (AEC). it means that competition in ASEAN countries increasingly open in all fields, one of them in the tourism sector. Tourism Industry is one of the country's revenue, especially in Indonesia. Indonesia tourism minister state that without a digital approach the tourism industry will die, because 70% of travellers already use online technology and therefore, as many as 50% conventional travel agents out of business [1]. The number of tourist arrivals to Indonesia has increased cumulatively in the period of January-November 2016 compared to the visit in January-November 2015, the number of foreign tourists visiting Indonesia reached 10.41 million visits or an increase of 10.46 percent compared to the number of foreign tourists visiting the same period in the previous year which amounted to 9.42 million visits [2].

Indonesia has many tourist attractions or objects. They contain nature, culture, and museum attractions. Some of nature attractions, for example, are *gunung bromo, danau toba, labuan bajo, mandalika, and tanjung kelayang*. Some culture attractions, for example, *kecak dance, salma dance, cultural ceremony*. Some museum attractions, for example, are *museum sangiran, museum angkut, museum national*. In this paper, we focus our investigation on indoor object and outdoor object, which contains museum attractions and nature attractions. The reason for the choice is that both object has destination classification, hence it is easier to filter from social media conversations.

Conversation in social media about general tourism, favourite destination, and one's perception play important role in a tourist's destination choice. The conversations contain abundant opinion perceptions, recommendations, and complains. By using network text, it is possible to summarize large-scale conversational data. The capabilities make the proposed method is more effective also efficient compared to natural text mining procedure such as *sentiment analysis* or *multi-class topic identification*. Network text detect frequency appeared words, then finding relationship between word, where one relationship means both words show up together in one sentence. The higher word association means they often appear together in many sentences. In this paper, the method is used to summarize dominant topics in complex conversational social network.

As a case study, we conduct experiments on two aspects of Indonesia tourism, namely indoor and outdoor tourist object. Our objective is specifically to extract public perception regarding museum and nature destination from *Twitter*. The proposed method of classifies topics and qualitatively explore any issues in each topic by giving meaning to each relation between words.

## 2. Theoretical Background

### 2.1. Consumer Behaviour and Perception

Consumer behaviour is the information about how consumers behave, how they make their decision about what they want or what they need regarding a product, service. In this paper context, it could be an organization process to see public choice about tourism object [3]. In this case, we would like to know how Indonesian tourist's behaviour and perception about tourism in Indonesia.

Perception is a process that begins with the consumer exposure and attention to how they feel, think attraction, and experience ends up with the interpretation about tourism object [4]. Each individual could have different perception about the same tourism object, therefore, we need to aggregate that perception information to our own benefit.

### 2.2. Marketing Intelligence

Marketing intelligence is a process to get and analyse information to understand the state of Indonesian's tourism and what can be potential opportunities in the future. The advantage of using marketing intelligence is that we can get information faster, efficient, and effective from external sources such as social media with the aim suits with the needs [5].

By using a marketing intelligence effort, we can see the competitor's condition, in this case, we can see how is the tourism perception in another country, thus we can learn how tourism in that particular country is going, hence we can make a better strategy for tourism in Indonesia.

### 2.3. Text Mining

Text mining is a discipline that combines data mining and text analytics using unstructured textual data. Along with structured data involves summarizing the content through text processing more efficient and effectively [6]. The most used application of text mining is sentiment analysis, with the objective to find the negative and positive sentiment towards products or services. Other than that, there is a multi-class identification to classifying data into one of the more than two categories.

In this research, we use text mining to summarize complex conversational data about tourism in Indonesia on social media. By using text mining, we can see the frequent appearance words to know the most talked object tourist. Combining with words association to see the relations between words, we can see the global pattern of topics in the social network. This methodology called *Network Text Analysis.*

*2.4. Social Network Analytics*

Social Network Analysis is a practical solution to model unstructured data in form of actors and their relationship. A network text methodology takes benefit from social network analysis construction [7]. The data source for social network analysis can be obtained from social media such as *Twitter. Twitter* provide open conversational data, this feature lead to high relationship data between actors in social networks. By transforming complex information about people thought and experience into relationship data, we can reduce drastically time needed to process information form social media. Thus, we say that this method is faster, easier, and cheaper

By using social network analysis, we recognize the relationship between tourist object and its perception. One of measurement in social network analysis is *Modularity*. This metric show how groups being formed into a network. *Modularity* is often used in optimization for detecting community structure in networks [8]. We use this metric to find out the topic group of each object tourist to see what the most talked regarding activities or attraction.

## 3. Methodology and Experimentation

We collect unstructured text data from social media *Twitter* by using a set of defined keywords concerning about tourism in Indonesia. The collection period is from *January 2017* to *April 2017*. As the result, we get in total *800.808* data tweets. From that data, we classify data into two groups, that is outdoor tourist object with 30.150 tweets and indoor tourist objects with 14.329 tweets. The workflow of network text analysis is shown in Fig 1.

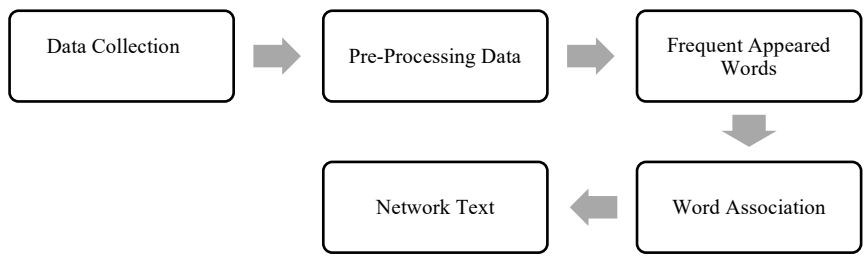

Figure 1. Workflow of network text analysis

The first step is data collection; we crawl twitter data directly about object tourist all around Indonesia. The second step, we filter the data using following procedures such as deleting irrelevant tweets or exclude the object that is not included in the topic we investigate. The third step is to find dominant words by using *the word cloud* applications. The fourth step is to calculate association rules between dominant words. The fifth step is to construct network text of dominant word, which includes weighted edge result for words association rules process. The last step, we analyse the result of the network, by qualitatively creating storytelling, context, and sense of network text. In network analysis, we employ centrality to find the most influential words in the networks and modularity to find words groups in the network. The data descriptions of each tourism object are shown in Table 1

TABLE I. *Data Profile of Each Tourist Object*

|  | **Indoor Tourist Object** | **Outdoor Tourist Object** |
|---|---|---|
| *Number of Tweets* | 14.392 | 30.150 |
| *After Filtering Process* | 4.952 | 16.864 |
| *Number of Word Groups* | 15 | 17 |

The results after words association rules process for both tourist object indoor and outdoor can be seen in Table II and Table III in the form of words pair list.

*TABLE II. Top 10 Indoor Words Pair*

| Words Pair | Weight Degree |
|---|---|
| museum_angkut-foto | 96 |
| museum_nasional-museum_gajah | 53 |
| museum_tekstil-batik | 35 |
| museum_sangiran-purba | 34 |
| museum_geologi-murah | 32 |
| museum_tsunami-aceh | 31 |
| museum_pos-bandung | 25 |
| museum_bank_indonesia-rekomnedasi | 24 |
| museum_ambarawa-foto | 22 |

*TABLE III. Top 10 Outdoor Words Pair*

| Words Pair | Weight Degree |
|---|---|
| Borobudur-Kuliner | 108 |
| Bromo-Sunrise | 76 |
| Seribu-Berlayar | 76 |
| Toba-Dayung | 75 |
| Wakatobi-Pelabuhan | 75 |
| Mandalika-homestay | 70 |
| Bajo-keindahan | 68 |
| Morotai-Murah | 54 |
| Lesung-Foto | 48 |
| Kelayang-Pantai | 40 |

## 4. Result and Analysis

For the case of indoor object shown in Fig. 2.a, we found that "*museum angkut*" is the most famous museum in Indonesia during data observation and in the words pair of "*museum angkut" and "foto*" is the most weighted connection, this refer to the most active and attraction in *museum angkut* is *foto* or in other words, the most popular attraction in *"museum angkot"* is taking pictures activity, or they have picturesque location. The second higher weighted connections word pair is "*museum nasional"* and *"museum gajah"* means that *museum nasional and museum gajah* is quite famous, with the indication that this object is often talked and visit. For the rest, we can see how words construct a together different perception about museums in Indonesia.

For the case of outdoor tourist object or nature object shown in Fig. 2.b, we found that "*candi borobudur"* is the most famous outdoor object, with the most dominant words pair "*candi borobudur" and "kuliner"* are the most weighted connection, this refer to the most activity and attraction in *candi borobudur* is *kuliner* or culinary activity. From that, we can explore and do more promote about culinary activity in *candi borobudur*. The second dominant topic is "*Gunung Bromo*" and "sunrise", this means that one of popular activity in *gunung bromo* is to see sunrise at the peak of *gunung bromo*. The fourth dominant topic is "*toba*" and "dayung" means that the most trending activity in *Danau Toba is "dayung"* or boating / rowing around the toba lake.

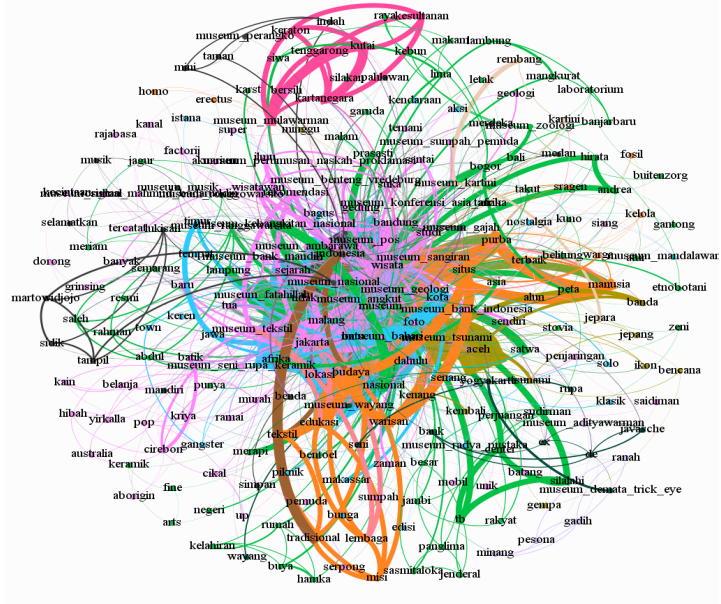

(a)

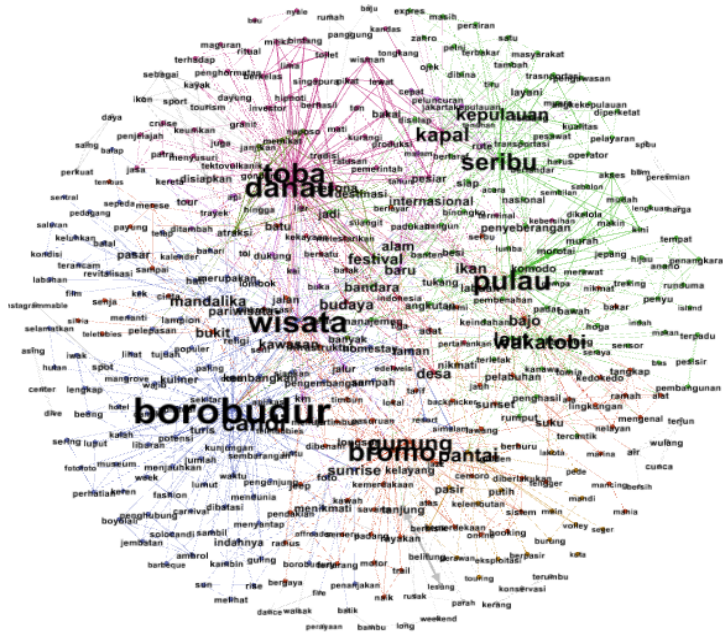

(b)

*Fig.2 Network text based in word pair and word cluster of (a) indoor object. (b). outdoor object*

From network text analysis methodology, we can see the activities and attractions on each tourist object. Furthermore, it also shows the interest of Indonesian people to choose and make the decision to visit tourist object what attract them the most. As the result, we can make a strategy to promote a tourism object. The resulting analysis of two tourist object both indoor and outdoor give us insight into perception about the tourism industry in Indonesia.

## 5. Conclusion

Network text analysis methodology proposed to help us to extract large-complex social network data from social media. We particularly mine the conversational data, which have more complex aspect than just review data, opinion data, or other one dimensional data. In the era where data is easy to gather, public need fast and cheap methodology to summarize and conclude those data. One particular application is in tourism industry, where people have overload information about any tourist object and other related information. Those information is also hard to verify from each source; thus, aggregation information mode is needed.

The proposed methodology can harness public perception, where it leads to marketing intelligence effort. Both end user or tourist and organizations, whether is business or government bodies can analyse to learn about how tourism in that country going and what the perception about tourism. The end objective for both is that we can a better strategy for each need.

As the conclusion, the proposed methodology proven than relevant summarizing content about tourism industry in more effective and efficient way. For further research, we propose to analyse another object from different social media platform to obtain a wider idea of the implementation.